\documentclass[%
 reprint,
superscriptaddress,
 amsmath,amssymb,
 aps,
prl,
floatfix,
]{revtex4-2}
\usepackage{amssymb}
\usepackage{mathrsfs}
\usepackage{graphicx}
\usepackage{dcolumn}
\usepackage{bm}
\usepackage{amsmath}
\usepackage{amsfonts}
\usepackage{color}
\usepackage{float}
\usepackage{makecell}  
\usepackage{multirow}  
\usepackage{threeparttable}  
\usepackage{booktabs}  
\usepackage{graphicx}
\usepackage{dcolumn}
\usepackage{bm}
\usepackage{hyperref}
\usepackage{varwidth}
\usepackage{subfig}
\usepackage{xcolor}
\hypersetup{
	colorlinks,
	linkcolor={red!50!black},
	citecolor={green!50!black},
	urlcolor={blue!80!black}
}
\usepackage{mathtools}

\DeclarePairedDelimiter\ket{\lvert}{\rangle}
\DeclarePairedDelimiterX\braket[2]{\langle}{\rangle}{#1\,\delimsize\vert\,\mathopen{}#2}
\makeatletter
\newcommand{\colorcaption}[2][]{%
  \begingroup%
  \renewcommand{\@caption@fignum@sep}{ (color online). }%
  \caption[#1]{#2}%
  \endgroup%
}
\makeatother

\begin{document}

\title{
Decoherence without einselection
}

\author{Xiao Zhang}
 \email{zhangxiao@mail.sysu.edu.cn}
\affiliation{School of Physics, Sun Yat-sen University, Guangzhou 510275, China}
\affiliation{Guangdong Provincial Key Laboratory of Magnetoelectric Physics and Devices, School of Physics, Sun Yat-sen University, Guangzhou 510275, China}  

\date{\today}
\begin{abstract}
Decoherence in a quantum measurement is typically explained as an interaction with the environment that destroys coherence between the system's eigenstates, a phenomenon known as environment-induced superselection (einselection). In this work, we demonstrate that einselection and the associated envariance are actually artifacts resulting from neglecting the non-equilibrium dynamics of the apparatus. We propose a new formalism of operator dressing, which we call the stochastic matrix integral (SMI), inspired by recent developments in quantum gravity algebras. This approach naturally arises from a modified Page-Wootters (PW) formula and describes decoherence as an interacting non-equilibrium process. It leads to the reduction of the Hilbert space and the emergence of an intrinsic non-unitary process as well as Born's rule. These outcomes are achieved without relying on the assumptions of einselection and pointer basis.
    
\end{abstract}

\maketitle
\noindent\textit{Introduction.--}
The incompatibility in quantum theory between reversible unitary evolution and irreversible wave function reduction during measurement \cite{rossi2021wigner}, along with the enigmatic nature of quantum gravity, are recognized by researchers as the two principal challenges in quantum theory \cite{cavalcanti2023fresh}. It has been suggested that these two puzzles may be correlated, either through quantum state reduction via the principle of equivalence or through fundamental time and length uncertainties induced by general relativity \cite{gambini2019single}.

However, although gravity provides compelling motivations to account for decoherence, or more formally, the measurement postulates or their equivalents in quantum theory \cite{masanes2019measurement}, none of these proposals address the fundamental question: why are only certain states corresponding to the eigenvalues of the observable stable and thus super-selected? Consequently, an ad hoc environment-induced superselection (einselection) hypothesis \cite{zurek2003} remains necessary, even when gravity is included in the decoherence picture \cite{gambini2019single}. In the following, we will discuss the issues with einselection and then show how very recent advancements in quantum gravity \cite{witten2022gravity, chandrasekaran2023algebra, leutheusser2023emergent} can help formulate a novel interpretation for decoherence without this hypothesis. We emphasize that we do not include gravity in the decoherence picture, but rather leverage recent developments in quantum algebras in gravity, such as the operator dressing \cite{chandrasekaran2023algebra} and Hilbert space reduction \cite{leutheusser2022subalgebra}.

Before diving into the application of recent developments in quantum gravity to decoherence, let us examine the einselection theory and its unsatisfactory aspects, which introduce a redundant "environment" into the "system-apparatus" combination, the minimal composition for performing a quantum measurement. In einselection, only certain composite "system-apparatus" pure states remain correlated, known as "pointer states" \cite{zurek2003}, and the coherence between the pointer states is destroyed, resulting in a classical mixture. Thus, einselection successfully describes decoherence as a non-unitary process involving information loss as mentioned earlier.

The forced involvement of the "environment" causes many problems. While the division between the "system" and the apparatus can be precise, it is difficult to separate the "apparatus" from the "environment," which are connected even before the measurement. Thus, how can we ensure that different ways of partitioning the "apparatus" and "environment" will result in the same selected states with probabilities untouched by Born's rule? Additionally, the byproduct of introducing the "environment" is "envariance" (environment-assisted invariance) as an emergent symmetry \cite{zurek2003}. This insists that the probability of all states is equal, which violates Born's rule. To accommodate Born's rule, a standard trick of "increasing resolution" used in classical probability must be applied, which lacks a physical origin, making it hard to imagine such a trick guaranteeing ultimate precision in quantum measurements. Moreover, how does the einselection mechanism automatically choose the correct set of pointer states for the apparatus, given that the sets of eigenstates of various operators are different?

Having reviewed the problems with einselection, let's explore how recent developments in quantum gravity can help remove these obstacles. CLPW\cite{chandrasekaran2023algebra} mentioned that an observer on a static path in de Sitter space converts the Type $\uppercase\expandafter{\romannumeral3}$ algebra to a Type $\uppercase\expandafter{\romannumeral2}_1$ algebra with operator dressing. Leutheusser and Liu have shown that, in the large N limit, the full Hilbert space of the boundary thermal field double state splits into disconnected GNS Hilbert spaces around semi-classical states via a Hawking-Page transition \cite{leutheusser2022subalgebra}. Inspired by these algebraic techniques, i.e. operator dressing and Hilbert space reduction, we developed a SMI and will demonstrate that it leads to decoherence without the need for einselection.

\noindent\textit{Decoherence with einselection.--}
Despite the insights of decoherence as a consequence of information loss through interaction, einselection theory \cite{zurek2003} has serious issues, as we have qualitatively stated in the introduction. Now we further present these issues in a mathematical format. According to einselection theory, we can write the resulting mixture of correlated states between the system $\ket{s_{j}}$ and the apparatus (observer) $\ket{O_j}$ selected by the environment basis $\ket{e_j}$ as follows:
\begin{equation}
\ket{\Phi_{SEO}}=\sum_{j=1}^{M}\ket{s_{j}}\ket{O_j}\ket{e_j},
\label{evariance}
\end{equation}
with the emergent envariance symmetry related to swapping \cite{zurek2003} and the consequent equal coefficients argument, the probability for each eigenvalue would be equal, i.e., $p(s_{j},O_j)=1/M$, which obviously violates Born's rule.

A standard trick of "coarse-graining" used in classical probability theory is applied in einselection theory to accommodate this. Born's rule is assumed to arise with the pointer basis of the apparatus (observer) $\ket{O_j}$ selected by the environment basis $\ket{e_j}$ through an interaction of c-shift \cite{zurek2003}. This can be written as:
\begin{equation}
\ket{\Phi_{SEO}}=\sum_{j=1}^{M}\ket{s_{k(j)}}\ket{O_j}\ket{e_j},
\label{local}
\end{equation}
where $S$ remains the same within "coarse-grained cells" indexed by $k$, and $j$ is introduced as a "fine-grained" index. Specifically, $M=\sum\limits^{N}_{k=1}m_k$, with $k(j)=1$ for $j\leq m_1$, $k(j)=2$ for $j\leq m_1+m_2$, and so on. Consequently, the probability of different Schmidt states of $S$ is $p(s_k)=m_k/M=|\alpha_k|^2$. This reproduces Born's rule.

So far, we can observe the serious problems that einselection introduces. Indeed, the "coarse-grained cells" lack a physical interpretation and suffer from the ambiguity of dividing the "apparatus" and the "environment." It is very hard to imagine that different divisions would result in precisely the same "apparatus-environment" and the same $m_k$ values, which Born's rule depends on. Moreover, it is mysterious how einselection can automatically choose the correct set of pointer states for different sets of eigenstates of various operators.
Indeed, Eq.\ref{local} represents an already finished state after decoherence. In the following, we will show that by considering decoherence as a dynamic process, the non-natural einselection assumption is naturally removed.

\noindent\textit{A modified PW formula and SMI.--}We start with the Schrödinger picture, where the operator $A=I_o\otimes A_s$ is stationary. Here, $A_s$ is a measurement operator acting on the system by the apparatus. We use the subscripts "s" and "o" to denote "system" and "observer" (apparatus), respectively. The density matrix $\rho_\tau$ describes the tensor product of the Hilbert space of the system and apparatus evolving with proper time $\tau$. Before we conduct the measurement, the system and apparatus are separate, and $\rho_\tau$ does not vary with time $\tau$. When we put them together, the interaction term $H_I$ is added to the Hamiltonian $H$ and they start to interact suddenly, as in a quantum quench \cite{mitra2018quantum}, and $[H,\rho_\tau] \neq 0$, so $\rho_\tau$ is no longer stationary. The conditional expectation value of $A$ is 
\begin{equation}
\begin{aligned}
E(A|\tau)=\frac{\mathop{tr}[AP_{\tau}\rho_\tau]}{\mathop{tr}[P_\tau\rho_\tau]},
\end{aligned}
\label{exp}
\end{equation}
where 
\begin{equation}
\begin{aligned}
P_\tau=I_s\otimes|\psi_o(\tau)\rangle\langle\psi_o(\tau)|.
\end{aligned}
\label{proj}
\end{equation}

Though Eq.\ref{exp} is identical to the starting point of the derivation of the PW formula \cite{page1983}, we will show that the consequent calculation deviates from it due to its non-equilibrium nature. Also note that what we are doing is different from relational quantum dynamics \cite{hohn2021trinity} since we are not quantizing time, but using absolute quantum mechanical time instead. We assume the total energy $H$ of the system and apparatus is conserved during the decoherence time, i.e., $H$ is the conserved charge associated with the Killing vector field in a semi-classical gravitational description \cite{chandrasekaran2023algebra}. However, the individual energy of the system and apparatus is not conserved and can vary with time, as long as their summation is conserved. Therefore, we can always write the total Hamiltonian effectively as $H=H_s(t)+H_o(t)$ \cite{Gemsheim2023}\footnote{The reason why we use $t$ instead of $\tau$ here is because we reserve $\tau$ as the integration upper bound from now on, which is manifest in Eq.\ref{exp_p1}}.

From Eq.\ref{exp} we have:
\begin{equation}
\begin{aligned}
&\mathop{tr}[AP_{\tau}\rho_\tau] \\
=&\mathop{tr}\left[A e^{-i\int_0^{\tau}(I_s\otimes [H_o(t)])dt}P_0 e^{i\int_0^{\tau}(I_s\otimes [H_o(t)])dt} \rho_\tau \right],
\end{aligned}
\label{exp_p1}
\end{equation}
Here we enclose $[H_o(t)]$ in square brackets to denote that it is a random (non-conserved) matrix Hamiltonian. Physically, this means that during a non-equilibrium interaction process, we cannot identify a certain portion of energy as belonging to the apparatus or system. The projection operator $P_o$ on the observer (apparatus) is dressed with an integral, describing its non-equilibrium time evolution, which we will show plays an essential role in interpreting decoherence without einselection.

The integration is over an stochastic trajectory of a non-conserved $H_o(t)$ over time $t$. It is not a path integral, as a path integral conserves the Hamiltonian. It resembles a matrix integral over random matrices in quantum gravity \cite{saad2019jt}, but the integration is not over matrices but over a possible time trajectory of them. It is a combination of stochastic calculus \cite{cohen2015stochastic} and random matrix integral, so we name it a stochastic matrix integral (SMI). We emphasize here that as long as the uncertainty in $H_s(t)$ is finite, no matter how small it is, the total number of possible trajectories $N$ always goes to infinity, which plays a key role in our analysis. This can be viewed as a peculiar quantum channel, which we will return to later.

We continue from Eq.\ref{exp_p1}. Due to $A$ commuting with $H_o(t)\otimes I_s$, we have:
\begin{equation}
\begin{aligned}
&\mathop{tr}[AP_{\tau}\rho_\tau] \\
=&\mathop{tr}\left[AP_0 e^{i\int_0^{\tau}(I_s\otimes [H_o(t)])dt} \rho_\tau e^{-i\int_0^{\tau}(I_s\otimes [H_o(t)])dt}\right] \\
=&\mathop{tr}\left[AP_0 e^{-i\int_0^{\tau}([H_s(t)]\otimes I_o)dt} \rho_0 e^{i\int_0^{\tau}([H_s(t)]\otimes I_o)dt}\right] \\
=&\mathop{tr}\left[e^{i\int_0^{\tau}([H_s(t)]\otimes I_o)dt} A e^{-i\int_0^{\tau}([H_s(t)]\otimes I_o)dt} P_0 \rho_0\right],
\end{aligned}
\label{exp_p2}
\end{equation}
The second equality is due to $\rho_\tau=e^{iH\tau}\rho_0 e^{-iH\tau}$.

Knowing that $\mathop{tr}[P_{\tau}\rho_\tau]=\mathop{tr}[P_0\rho_0]$, the expectation value of the operator $A$ can be written as:
\begin{equation}
\begin{aligned}
E(A|\tau)=\mathop{tr}\limits_{s}[A_s(\tau)\rho_{s0}],
\end{aligned}
\label{PW}
\end{equation}
where $A_s(\tau)$ is dressed by an SMI as $e^{i\int_0^{\tau}[H_o(t)]dt}A_s e^{-i\int_0^{\tau}[H_o(t)]dt}$. $\rho_{s0}$ is the relative density matrix at time $0$:
\begin{equation}
\begin{aligned}
\rho_{s0}=\frac{\mathop{tr}\limits_{o}[P_0\rho_0]}{\mathop{tr}[P_0\rho_0]}.
\end{aligned}
\label{dmatrix}
\end{equation}
Because the apparatus and the system are initially separate $\rho_0=\rho_{s0}\otimes\rho_{o0}$, we have:
\begin{equation}
\begin{aligned}
\rho_{s0}=\frac{\mathop{tr}\limits_{o}[P_0\rho_0]}{\mathop{tr}[P_0\rho_0]}=\frac{\rho_{s0}}{\mathop{tr}[\rho_{s0}]}.
\end{aligned}
\label{dmatrix_simple}
\end{equation}
In our Eq.\ref{PW}, the SMI dressing of $A_s$ is different from the Heisenberg-evolved operator in the PW formula $A_s(\tau)=e^{iH_s\tau}A_s e^{-iH_s\tau}$ \cite{page1983}. Note that $A_s(0)=A_s$, and in the following, we will just use $A_s$ to avoid confusion.

\noindent\textit{SMI without pointer states.--}It is obvious that the original Heisenberg evolution of operators in the PW formula cannot lead to decoherence. Before we show how our SMI can achieve this, let us further explore its properties. The SMI describes a non-equilibrium process with non-conserved energy, involving a stochastic integral over one of the (infinitely) possible paths of the non-conserved random matrix Hamiltonian $H_s(t)$. We can write it as a unique quantum channel $\mathcal{N}_{s\otimes o\rightarrow s}$\cite{wilde2011classical}:

\begin{equation}
A_s(\tau)=\mathcal{N}_{s\otimes o\rightarrow s}(A_s)=\mathcal{K}(\tau)A_s \mathcal{K}^{\dag}(\tau),
\end{equation}
where

\begin{equation}
\mathcal{K}(\tau)=e^{i\int_0^{\tau} [H_s(t)] dt}.
\end{equation}

It is important to note that $[H_s(t)]$ usually does not commute with the observable $A_s$ at any time $t$, i.e., $[A_s(t),\mathcal{K}(t)] \neq 0$, and the quantum channel is not a completely positive and trace-preserving (CPTP) quantum channel, thus it does not preserve unitarity.

\begin{figure}[htbp]

    \subfloat[]{
	\centering
	\includegraphics[scale=0.27]{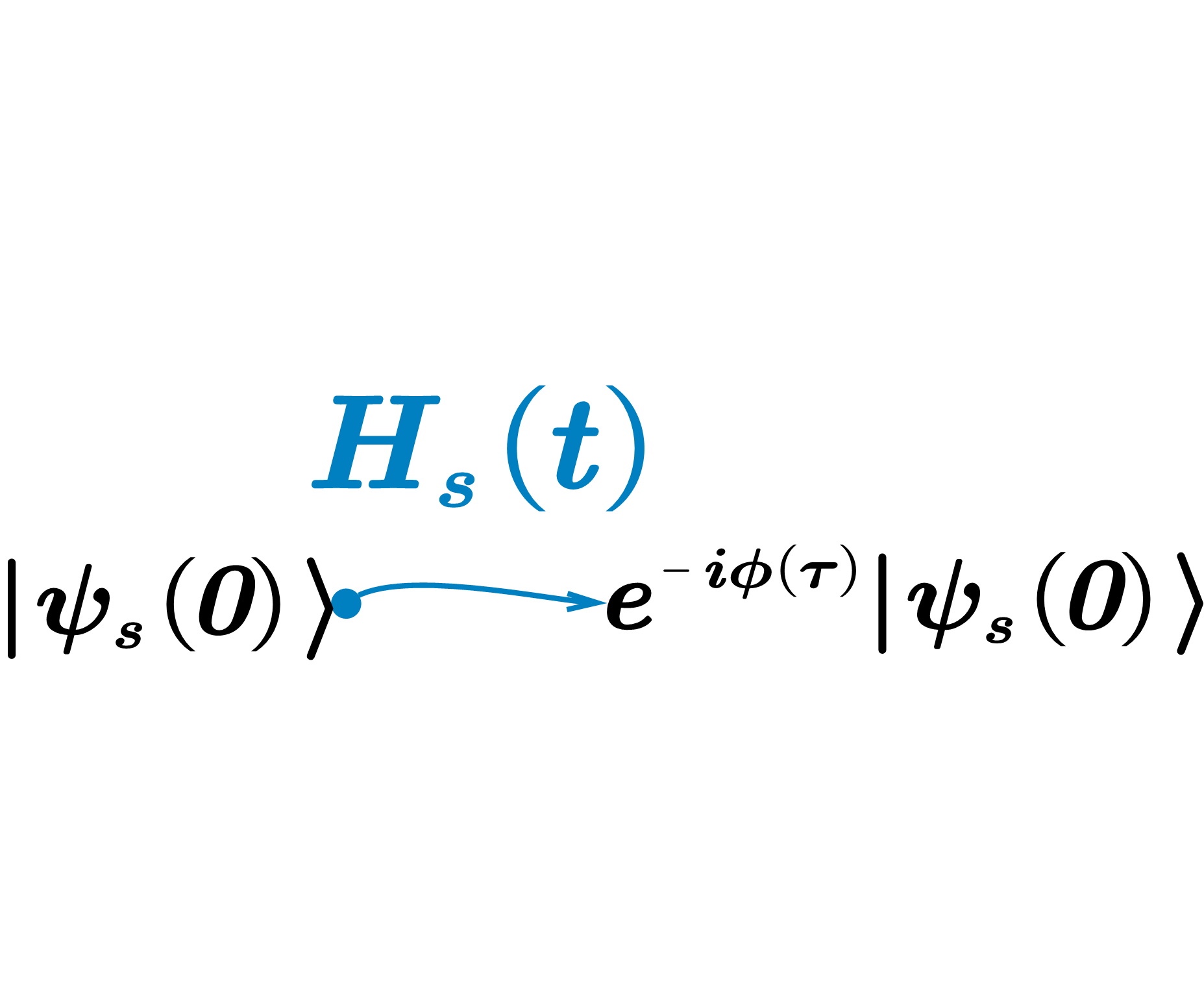}
    }
    \subfloat[]{
        \centering
	\includegraphics[scale=0.27]{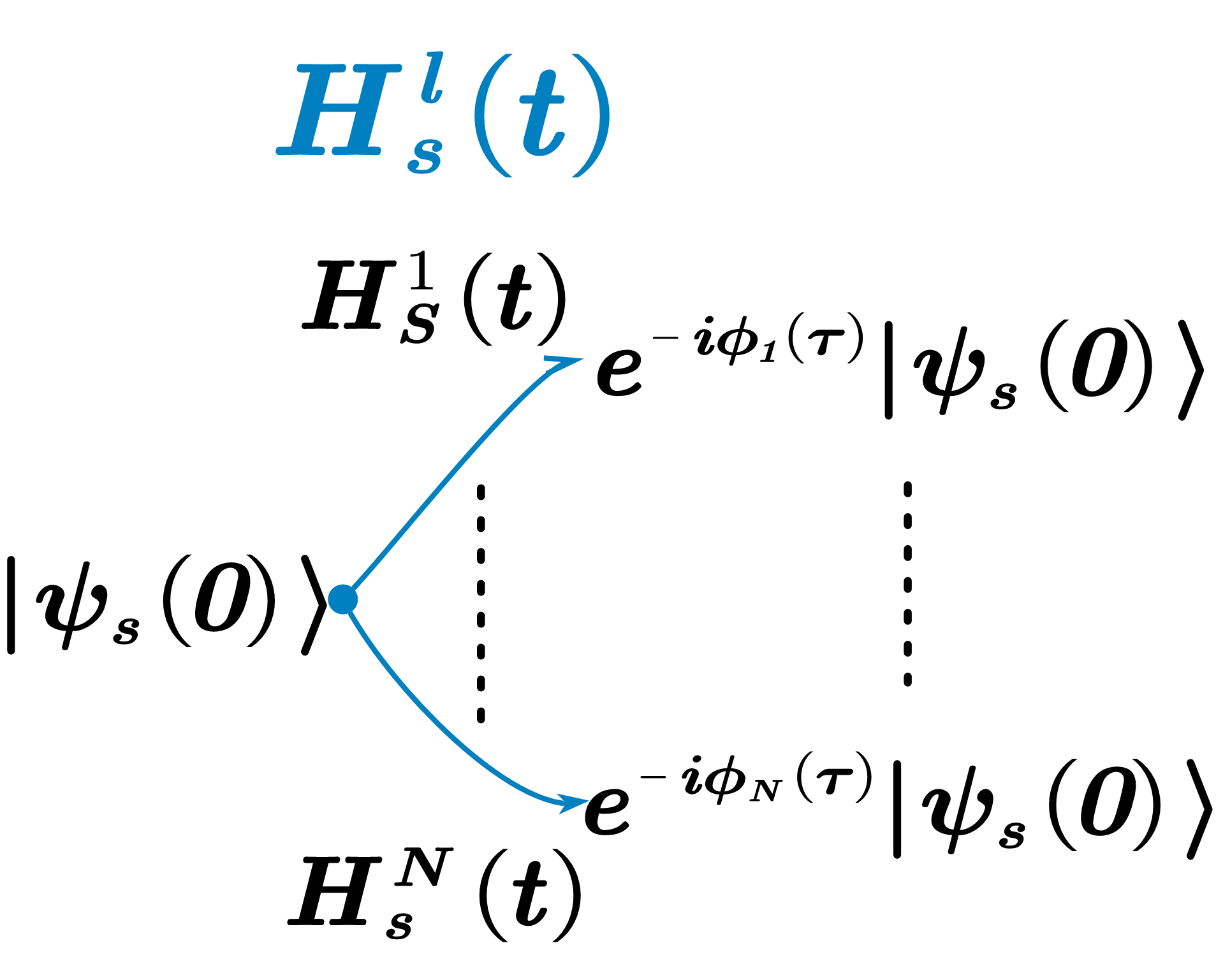}
    }
    
	\caption{The evolution of system wavefunctions: (a) projected onto a quantum clock or pointer state as a reference state, resulting in a conserved system Hamiltonian $H_s(t)$ and a fixed trajectory; (b) under SMI with a non-conserved system Hamiltonian $H_s(t)$, resulting in infinite possible trajectories ($N\rightarrow \infty$). Each trajectory is labeled by $l$.}
\label{evolution} 
\end{figure}
The difference between our SMI and the previously discussed equilibrium system evolution under a quantum clock \cite{smith2019quantizing} or pointer state \cite{Gemsheim2023} is shown in Fig.~\ref{evolution}. In the case of either a quantum clock or pointer state shown in Fig.~\ref{evolution}(a), the time evolution of $H_{o/c}(t)$ at any time $t$ is fixed, resulting in a fixed evolution of the system under $H_s(t)=H-H_{o/c}(t)$. Note that the subscript $o/c$ in $H_{o/c}(t)$ indicates apparatus pointer/clock state. The pointer states measure the system while the clock state measures time, but they all result in a fixed system evolution. 

When we consider decoherence as a non-equilibrium dynamic process, only the total energy $H$ is conserved and $H_s(t)$ varies under different evolution trajectories of the system as shown in Fig.~\ref{evolution}(b). This "blurring" in trajectories can be understood with reference to Eq.~\ref{PW}, where the information regarding the dynamics of the apparatus is lost\footnote{The apparatus does not measure itself.}, which can be viewed identically as the information uncertainty of the system. We will show that it is indeed the non-existence of pointer states as a fixed reference state that leads to decoherence, contrary to common thinking!

\noindent\textit{Reduction of Hilbert space through SMI.--}A process that can change the algebra and underlying Hilbert space is usually ignored in ordinary quantum mechanics, where the von Neumann algebra is implicitly assumed to be Type $I$ \cite{sorce2023notes}. Inspired by the very recent developments of operator dressing in quantum gravity \cite{chandrasekaran2023algebra}, we prove that the SMI dressing, as an integral over any possible trajectory of a non-conserved system Hamiltonian $H_s(t)$, naturally reduces the Hilbert space into a classical mixture and reduces $A_s$ to Born's rule $A_s=\sum_{i}a_iP_{a_i}$, the so-called decoherence.

\begin{figure}[htbp]
    \subfloat[]{
	\centering
	\includegraphics[scale=0.24]{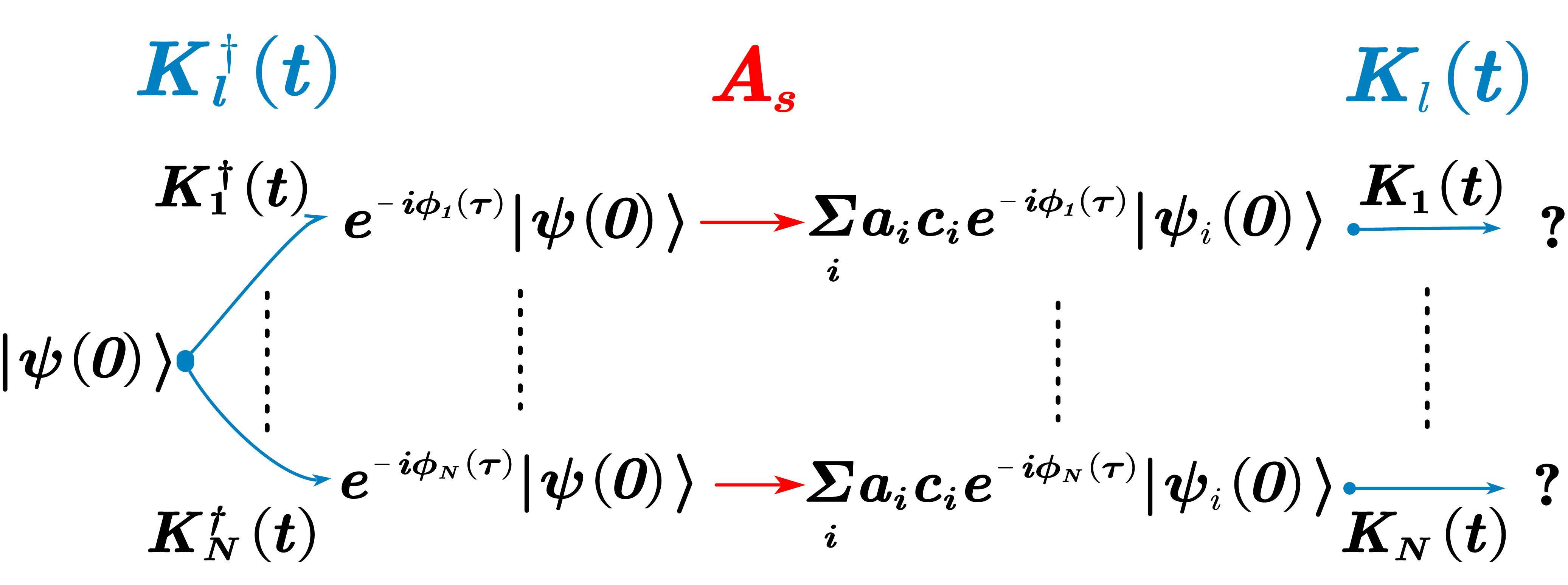}
    }\\
    \subfloat[]{
        \centering
	\includegraphics[scale=0.24]{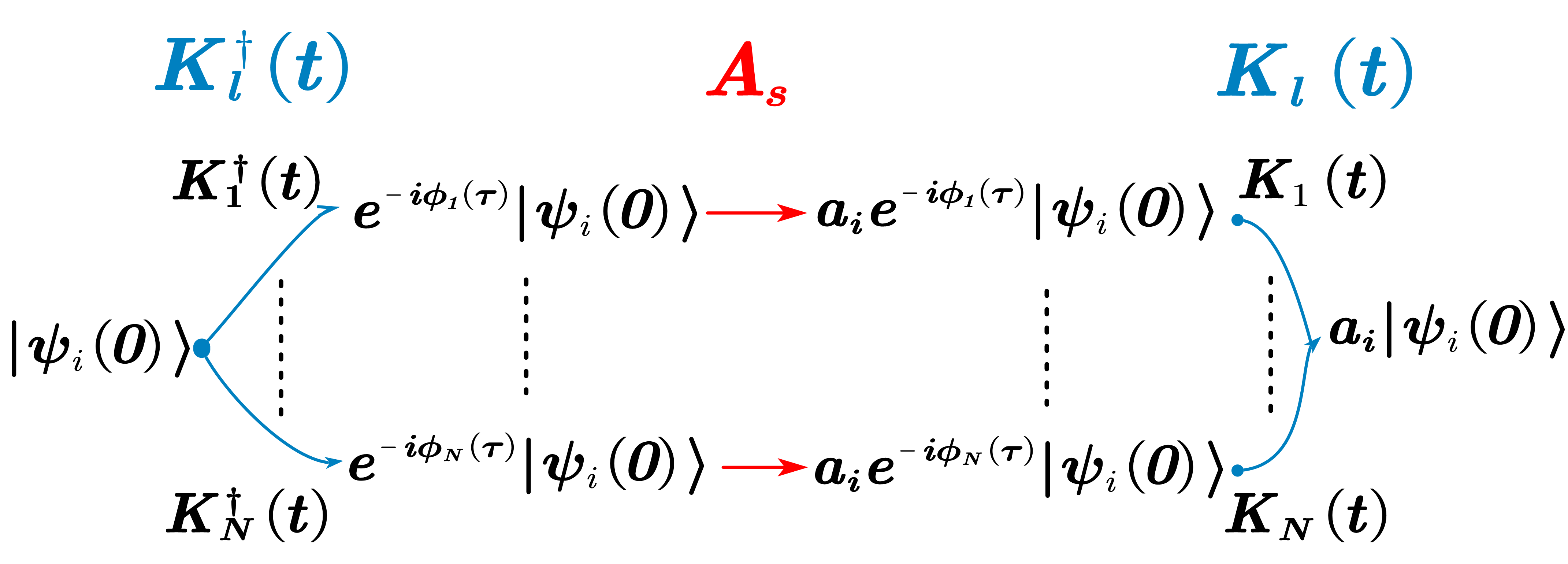}
    }
	\caption{Stability criteria for observable system states. Note that in these figures we omit the subscript "s" for better visualization. The operator with SMI dressing is (a) not well-defined on a non-eigenstate $|\psi_s(0)\rangle$, while it is well-defined on (b) its eigenstate $|\psi_{s_i}(0)\rangle$.  The difference between (a) and (b) is that when a certain state evolves under $\mathcal{K}_l^{\dag}$ and then interacts with $A_s$, if it is not an eigenstate, it will split into a few eigenstates $e^{-i\phi_l(\tau)}|\psi_{s_i}(0)\rangle$ with coefficients $c_i$ multiplied by individual eigenvalues $a_i$. It will then be scattered by the quantum channel operator $\mathcal{K}_l$ and can no longer return to the original pure state.}
\label{stability} 
\end{figure}

To begin with, a system state $|\psi_s(0)\rangle$ evolves first under the channel operator $\mathcal{K}_l^{\dag}$ of the $l_{th}$ trajectory, as shown in Fig.~\ref{stability}a, in which we omit the subscript "s" for better visualization. As a ground state of each channel, it will acquire a different phase $e^{-i\phi_l(\tau)}$ through adiabatic evolution depending on $l$ and a varying $\tau$. The adiabatic assumption is completely understandable since a measurement should be perturbative. Then, when it interacts with $A_s$, if it is not an eigenstate, it will split into a few eigenstates $e^{-i\phi_l(\tau)}|\psi_{s_i}(0)\rangle$ with coefficients $c_i$ multiplied by individual eigenvalues $a_i$. Subsequently, it will be scattered by the time reversal operator $\mathcal{K}_l$ of $\mathcal{K}_l^{\dag}$, and can no longer return to the original pure state. In contrast, as shown in Fig.~\ref{stability}b, an eigenstate $|\psi_{s_i}(0)\rangle$ of $A_s$ can be observed, as it always remains stable after being acted upon by $A_s(\tau)$ while the apparatus measures an eigenvalue $a_i$.

We emphasize that $A_s(\tau)$ is only well defined when all the infinite possible trajectories resolve to the same result, in which any non-eigenstate will become undefined when acted upon by it, thus becoming unstable and dropping out. An alternative way of saying this is that the measurement operator cannot take eigenstates to ones with different eigenvalues when SMI dressing is applied. This is somewhat similar to the simulation in the quantum gravity process mentioned in the introduction: all the operators that take states from one disjoint "continent" of Hilbert space to another do not survive in the large $N$ limit, which separates the Hilbert space into disjoint "continents" of Hilbert spaces \cite{leutheusser2022subalgebra}.

In this way, $A_s(\tau)$ will project any initial wavefunction of the system $|\psi_{s}(0)\rangle$ into a mixture of eigenstates of $A_s$ as $A_s(\tau)=\sum_{i}a_iP_{a_i}$ while all the other states drop out. Note that in the above derivation we have implicitly assumed a \textit{no super-selection rule}: every eigenstate has an equal probability of being mapped to by $P_{a_i}$. It is a natural assumption which indicates that no eigenstate has priority over another during a measurement.

\noindent\textit{Conclusion and outlook.--}In the above analysis, we have demonstrated that when a measurable operator is dressed by an SMI, Born's rule emerges with a reduced Hilbert space as a classical mixture of eigenstates of the operator without einselection. Our algebra remains Type I since we still have projection operators with integer rank \cite{sorce2023notes}, but with a reduced Hilbert space. Even if the total energy $H(t)$ of the system-apparatus combination is not conserved and has a time $t$ dependence, the entire derivation and conclusion still hold.

It is worth mentioning that we have been cavalier about the magnitude of $[H_s(t)]$ and $\tau$. First, in our derivation, $H_s(t)$ could vary from $0$ to $H$ regardless of the interaction's magnitude. However, in the limit $H_I \rightarrow 0$, Eq.~\ref{exp_p1} should reduce to the PW formula, with a conserved $H_s$ that does not evolve with time, without ambiguity. This suggests that the configuration space in which $H_s(t)$ could vary should be restricted by the interaction's strength, a relationship that has yet to be derived. Another unrestricted quantity in our derivation is the decoherence time $\tau$. It has been suggested that a strong interaction would require a shorter decoherence time scale \cite{gambini2019single}, which awaits further investigation to understand its physical consequences in SMI.

Finally, we would like to point out the connection between our work and recent developments in quantum gravity. It has been found that the two-dimensional Jackiw-Teitelboim (JT) gravity can be evaluated as a matrix ensemble \cite{saad2019jt}. Considering that our SMI can also reduce a pure state into an ensemble, it would be interesting to explore this in a quantum gravity context. We will leave this for future work.

\textit{Acknowledgements--}
X.Z. is supported by the National Natural Science Foundation of China (Grant No. 11874431), the National Key R\&D Program of China (Grant No. 2018YFA0306800). The calculations reported were performed on resources provided by the Guangdong Provincial Key Laboratory of Magnetoelectric Physics and Devices, No. 2022B1212010008.

%


\end{document}